*The Transactional Interpretation and its Evolution into the 21st Century: An Overview*
*R. E. Kastner*
*July 28, 2016*

> *"Now in the further development of science, we want more than just a formula. First we have an observation, then we have numbers that we measure, then we have a law which summarizes all the numbers. But the real glory of science is that we can find a way of thinking such that the law is evident." -Richard Feynman, The Feynman Lectures on Physics, Vol. 1*

ABSTRACT. This essay provides a historical, philosophical, and critical overview of the development of the Transactional Interpretation of Quantum Mechanics (TI). It is separated into two parts. Part I presents the history and development of TI from 1986 up to 2016 (the time of writing). Part II lays out current areas of divergence among researchers in TI.

Part I

1. Historical Background

The transactional interpretation of quantum mechanics (TI) was initially formulated by John G. Cramer (Cramer, 1986). The status of TI as an interpretation is an interesting one: at the non-relativistic level, it is not really a different theory from standard quantum mechanics, and can be considered simply an alternative interpretation. However, at the relativistic level it is actually a different theory from quantum field theories such as quantum electrodynamics (QED), since it is based on a direct-action theory of fields rather than on the usual quantized fields. This has become evident as a result of its relativistic development (Kastner 2012, 2014a, 2015), which is based on the direct-action theory of Davies (1971, 1972).

TI was inspired by the Wheeler-Feynman (WF) time-symmetric or 'direct action' theory of classical electrodynamics (Wheeler and Feynman 1945, 1949). In the WF theory, radiation begins as a time-symmetric process. A charge emits a field in the form of half-retarded, half-advanced solutions to the wave equation; the half-retarded, half-advanced response of absorbers then gives rise to a radiative process that transfers energy from an emitter to an absorber. Cramer applied this picture to quantum theory, proposing that quantum states were emitted in a half-retarded, half-advanced processes and that absorbers generated responses, just as in the Wheeler-Feynman theory. In this picture, the usual quantum state is called an 'offer wave' (OW) and the advanced response from the absorber is called the 'confirmation wave' (CW). In terms of state vectors, the OW is represented by a ket, $|\Psi>$, and the CW by a dual state vector or 'brac,' $<\Phi|$. [1] Cramer coined the term

---

[1] While the standard term is 'bra', we use 'brac' here since it seems more consistent with the term 'bracket notation', given that the usual state is called a 'ket'.

'transaction' for this two-way process between emitters and absorbers, because it is analogous to a financial transaction.

2 How TI explains von Neumann's Measurement Process and the Born Rule

A key feature of TI is that it provides a physical basis for the quantum probability rule (the Born Rule) for the outcomes of measurements. In this regard, it can be seen as performing the explanatory function alluded to by Feynman in the quoted epigram above. The Born Rule, initially arrived at through an ansatz by Max Born and found to be empirically valid, specifies that the probability of an outcome $x$ is given by the absolute square of the wave function $\Psi(x)$. The natural origin of the Born Rule from TI will be discussed in detail below.

John von Neumann (1955) delineated two different processes that take place in quantum systems. The second of these, which he called 'Process 2', is the ordinary unitary evolution described by the Schrödinger Equation. The first of these, called 'Process 1', is the transition of a quantum system from a pure state to a mixed state upon measurement, i.e.:

$$|\psi\rangle \to \sum_n |c_n|^2 |\psi_n\rangle\langle\psi_n|$$

$$(1)$$

The coefficients $|c_n|^2$ are the probabilities given by the Born Rule for each of the outcomes $\psi_n$. For clarity, we should distinguish two different stages of Process 1. The first (a) corresponds to the transition in (1) above; the second (b) is the 'collapse' in which one outcome is actualized, corresponding to the choice of a particular outcome $n=k$.

Von Neumann noted that this Process 1 transformation is acausal, nonunitary, and irreversible, yet he was unable to explain it in physical terms. He himself spoke of this transition as dependent on an observing consciousness. However, one need not view the measurement process as observer-dependent. If the advanced responses of absorbers are taken into account, then for an OW described by $|\Psi\rangle$, we have for a collection of numbered absorbers:

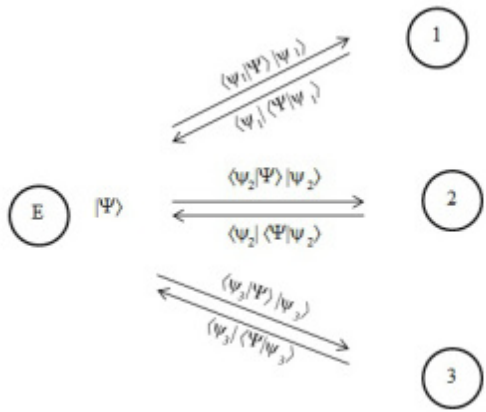

 In the above diagram, an initial offer wave from emitter E interacts with some measuring apparatus that separates it into components $\langle\psi_n|\Psi\rangle\ |\psi_n\rangle$, each reaching a different absorber $n$. Each absorber responds with an advanced (adjoint) confirmation $\langle\ \psi_n\ |<\Psi\ |\ \psi_n\ \rangle$. Cramer showed in his (1986) that the final amplitude of the confirmation from each absorber, upon reaching the emitter, is $\langle\psi_n|\Psi\rangle<\Psi\ |\ \psi_n\ \rangle$. Thus, each emitter/absorber interaction is described by the product of the OW and CW, which gives a weighted projection operator: $\langle\psi_n|\Psi\rangle<\Psi\ |\ \psi_n\ \rangle\ |\ \psi_n\rangle<\psi_n\ | = |c_n|^2\ |\ \psi_n\rangle<\psi_n\ |$. In TI, these OW/CW encounters are called *incipient transactions*. If we add all the incipient transactions, we clearly have the density operator representation of 'Process 1".[2]

Thus, by including the advanced responses of absorbers, TI yields a physical account of the Von Neumann measurement transition ("Process 1"), as well as a natural explanation of the Born Rule. The response of absorbers is what creates the irreversible act of measurement and breaks the linearity of the basic deterministic propagation of the quantum state. Since the conserved physical quantities can only be delivered to one absorber, there is an indeterministic collapse into one of the outcomes $\psi_k$ with a probability given by the weight $|c_k|^2$ of the associated projection operator $|\ \psi_k\rangle<\psi_k\ |$. The latter corresponds to stage (b) of the measurement transition. This is called an *actualized transaction*, and it consists in the delivery of energy, momentum, angular momentum, etc., to absorber $k$. That absorber that figuratively wins the incipient transaction 'lottery' is called the 'receiving absorber' in PTI. The process of collapse precipitated in this way by absorber response(s) can be understood as a form of spontaneous symmetry breaking (this is discussed in Chapter 4 of Kastner 2012).

3. Relativistic, possibilist extension of TI

---

[2] It is already evident at this point that such a process, if taken as physically real, cannot be something that occurs entirely within a single spacetime, in which only one outcome ever occurs for any measurement. This issue is addressed in the next section.

The basic transactional picture has been extended by Kastner, in a possibilist ontology, to the relativistic domain. This version of the interpretation is referred to as 'Possibilist Transactional Interpretation' or PTI (Kastner 2012). A theoretical basis can be found in the Davies application of the direct-action picture of fields to quantum electrodynamics (Davies 1971, 1972).[3] However, PTI departs from the Davies treatment in two ways: (i) virtual particles are clearly distinguished from real particles (see also Kastner 2014a) and (ii) the coupling amplitude, in this case the elementary unit of charge *e*, is identified as the amplitude for generation of an offer or confirmation wave, in a transactional account.[4] These developments allow the interpretation to provide a smooth transition between the non-relativistic and the relativistic domains. The latter can be viewed as the 'birthplace' of offer waves, in a physically well-defined (although fundamentally stochastic) manner. This process will be described in Section 4 below.

The ontology must be a possibilist one -- i.e., one in which the quantum possibilities, such as offer waves, are considered as real -- because of the mathematical structure of quantum theory, as follows. If the system under consideration consists only of a single quantum, its associated state vector has only 3 spatial degrees of freedom. In that case, one can think of the wave function as inhabiting spacetime. However, any composite system of N quanta has 3N spatial degrees of freedom, and therefore, cannot be considered a spacetime object. A realist interpretation of the quantum state must take this fact into account.

If we think of the spacetime realm as the realm of concrete, actualized events, then the quantum entities described by state vectors must have a different ontological status. In PTI they are viewed as physical possibilities or *potentiae*, just as Heisenberg suggested.[5] The latter are precursors to any actualized spacetime event. In view of quantum indeterminism, they are necessary but not sufficient conditions for observable events. The necessary condition is the emission of an offer wave and confirming response(s) to that offer wave from one or more absorbers. This sets up a set of incipient transactions as described in the previous section. At that point we have non-unitary collapse, and only one of the set is actualized to become a spacetime interval with well-defined emission and absorption endpoints—that is the sufficient condition. Thus, not all OW/CW exchanges will result in actualized transactions corresponding to spacetime events.

## 4 The relativistic domain as the birthplace of offer waves

Nonrelativistic quantum mechanics describes a constant number of particles emitted at some locus and absorbed at another. However, in the relativistic domain we are dealing with interactions among various coupling fields, and the number and types of quanta are generally in great flux. Such interactions are described in terms of scattering. The internal connections –i.e., virtual particles-- are characterized at each scattering vertex by the coupling amplitude (in the case of QED, the elementary charge $e$) and are neither offers nor confirmations according to relativistic PTI. These are 'internal lines' in which the direction of propagation is undefined; i.e., there is no fact of the matter concerning which current 'emitted' and which 'absorbed' the virtual quantum. In that case, as shown by Davies (1971), the Feynman propagator $D_F$ of standard QED can be replaced by the time-symmetric propagator in the direct-action theory.

---

[3] It should be noted that the direct-action theory of fields provides an elegant and effective escape from Haag's Theorem, a famously vexing result showing that the interaction picture of the standard quantized fields does not exist. This point is discussed in Kastner (2015).

[4] Here *e* is expressed in natural units such that it becomes a dimensionless number, the square root of the fine structure constant $\alpha$ of quantum electrodynamics. PTI can in principle be generalized to forms of charge, such as the color charge.

[5] Heisenberg (2007), p. 15

According to PTI, the field coupling amplitudes, which are not present in the non-relativistic case, represent *the amplitude for an offer or confirmation to be generated*. This is a feature of the interpretation appearing only at the relativistic level, in which the number and type of particles can change. It is a natural step, in view of the fact that in standard QED the coupling amplitude is the amplitude for a real photon to be emitted. In PTI, a 'real photon' corresponds to an offer wave (and its set of confirming responses, see also below). So we can think of virtual particles as necessary but not sufficient conditions for real particles, which correspond to transactions in PTI. We see that the relativistic level brings with it a subtler form of the uncertainty associated with the nonrelativistic form of the theory, the latter being the amplitudes of the offer wave components reaching various absorbers.

Thus, in PTI there is a clear distinction between virtual and real particles: virtual particles are not offer waves. Rather, they are *precursors* to offers or confirmations that do not rise to that level. In general, they are not on the mass shell. In the direct-action theory underlying the transactional picture, they are represented by time-symmetric propagators, not Fock space states. On the other hand, real particles correspond to offer waves, which are on the mass shell and are represented by Fock space states. In the direct action picture, offer waves are always responded to by confirmations, so the term 'real photon' can refer either to a Fock space state |k> (i.e., just to the OW component) or to the projection operator |k><k| depending on the context.

One of the criticisms of the original Cramer theory was that it took emitters and absorbers as primitive. PTI overcomes this limitation by providing well-defined physical conditions for the generation of OW and CW: both have their origins in the incessant virtual particle activity that is the coupling between fields, such as between the Dirac field and the electromagnetic field. When the interacting photons are on the mass shell and satisfy energy conservation, they are capable of achieving OW status, which generates a CW response from eligible interacting fermionic currents. This is an inherently indeterministic process, and is described by two factors of the coupling amplitude—i.e., the fine structure constant for QED (as well as the square of the relevant transition amplitude between the initial and final states, with satisfaction of energy conservation). The two factors of the coupling amplitude correspond to the two vertices that are involved in each scattering interaction.

Details of the conditions for the emission of an offer wave, as opposed to an internal process (i.e. a propagator linking two vertices), are given in Kastner (2014a), along with a discussion of how this picture can shed light on the computations necessary to obtain atomic decay rates. Decays are spontaneous emission processes in which virtual photon exchange is spontaneously elevated to a real photon offer and confirmation. Such elevations must of course satisfy energy conservation. That is the only way a would-be virtual photon becomes a real photon—only those photons satisfying the mass shell condition and the energy conservation requirements are eligible to be elevated in this way. The account of antimatter in PTI is similar to that of Feynman's, despite the fact that the time-symmetric propagator replaces the Feynman propagator at the virtual particle level (details are given in Kastner 2016).

## 5  Spacetime as the growing set of actualized events

It is often claimed that relativity implies a block world – that is, an ever-present spacetime in which all past, present and future events exist in an equally robust sense. The main argument used in support of that claim is termed 'chronogeometrical fatalism' (cf. Stein 1991, pp. 148-9).  However, that argument rests on certain assumptions, such as the ontological status of 'lines of simultaneity', and a substantival view of spacetime as a 'container' for events, that do not necessarily hold. See Kastner 2012, §8.1.4 for a rebuttal of chronogeometrical fatalism. See also Sorkin (2007) for a rebuttal to this common but erroneous assumption that a 'block world' is necessarily implied by relativity.

A different model, of a growing spacetime, is perfectly viable; one such model is the 'causal set' approach as proposed by Bombelli et al (1987) (see also Sorkin (2003) and Marolf and Sorkin (2006)).  Recently, Knuth and his collaborators have provided an information-based account of the causal set spacetime ontology based on influences between observers, in which common observer rest frames, the Minkowski metric, and even a notion of geodesic curvature naturally emerge (cf. Knuth and Bahreyni (2014), Walsh and Knuth (2015).  Since in PTI, all such influences are transactions, the transactional process can readily serve as the dynamics generating the growth of this kind of covariant spacetime causet: the probabilities of occurrence of transactions are Poissonian (a requirement for relativistic covariance of the set). This is because the generation of OW and their CW responses are described by decay rates.

## Part II Diverging Viewpoints on TI and Critical Assessment

## 1. The Maudlin Challenge

TI was subject to a challenge by Maudlin (1996) who argued that it gave rise to inconsistencies and causal loop problems based on situations with moveable absorbers whose availability to the emitter is contingent on a particular detected outcome. These experiments were termed "contingent absorber experiments" by Kastner (2012, Section 5.1). Maudlin assumed that a slow-moving OW is emitted, but one of the detectors is only made accessible to it if another detector fails to detect the particle. He argued that this resulted in an inconsistency concerning the existence of a confirmation from side with the moveable detector, as well as inconsistency in the assigned probabilities. This issue has been addressed and resolved by Marchildon (2006) and Kastner (2006, 2012). Marchildon pointed out that in the absorber theory, there is always complete absorption of any emitted offer, so there must be a background absorber on the side originally missing the moveable detector, which always returns a CW.

The Maudlin challenge is no longer viewed as a fatal problem, although it prompted Kastner to further develop the interpretation to articulate the ontological issues in play in contingent absorber experiments. In particular it is observed in Kastner (2012), Section 5.1 that similar apparent paradoxes appear even in standard quantum mechanics in situations involving delayed choice. Thus these apparent paradoxes are not unique threats to TI but are features inherent in standard nonrelativistic quantum theory.

However, when considering the relativistic details of the PTI model, one finds that the Maudlin challenge cannot even be mounted. For one thing, a bound state such as a slow-moving atom is not an OW (see Kastner 2016). Only confirmed field excitations are OW. Thus, in order to realize the Maudlin experiment, one would need to use a field excitation with nonvanishing rest mass, such as an electron, as the putative 'slow-moving particle.' However, even this would not instantiate the proposed experiment, since quasi-free electrons[6] (such as those generated by beta decay, photoelectric effect, or other such processes) do not generate their own CW. They participate in transactions through their coupling with the electromagnetic field (through the fields generated by charged currents), so it is the attendant photons that generate the CW, not the electrons. Further details as to why the Maudlin experiment cannot be instantiated in TI are being discussed in a forthcoming work. These details involve consideration of the relativistic level of the relevant processes, which Maudlin did not take into account in his proposal.[7]

2. Cramer's 3+1 spacetime ontology

In his original presentation of TI (Cramer, 1986), Cramer used the term 'pseudotime' as the domain in which dynamical OW/CW interactions took place, but stressed that it was just a heuristic tool, and did not correspond to any real physical domain. It is evident from his recent account (Cramer, 2016) that he wishes to keep the ontology strictly in 3+1 spacetime. However, if a single spacetime is the extent of the ontology, there does not appear to be any room for the OW/CW interactions and the choice of one of many incipient transactions resulting in collapse and actualization of the associated events. Along the same lines, there is no room in the ontology for the dynamical hierarchical account of transactions that Cramer presents as a rebuttal to the Maudlin challenge (which is not necessary given the relativistic developments discussed above). Such dynamical stories do not have anything to refer to in a single 3+1 spacetime.

---

[6] The term 'quasi-free' is used here, since such electrons are actually transitioning from one bound state to another. No free electron can emit or absorb a photon due to energy conservation.

[7] For example, the inverse of beta decay, electron capture, only occurs for a bound electron OW that is already part of an atom, not a quasi-free external electron OW. Thus, a pure electron transaction can only occur between a bound electron OW and its atomic nucleus. An electron OW could be incorporated into a cation (atomic ion resulting from loss of an electron from its outer shell), thus de-ionizing it, but in that case no CW is generated. (See Kastner 2016 for a discussion of bound states, such as atoms, as conglomerates of unconfirmed offer waves). Also, in electron scattering with photons (which is how we detect electrons), it is the photon confirmation that actualizes the electron. No electron confirmation need occur in this process. This is because of the natural asymmetry between charged currents and photons; the former are sources of the latter, but not vice versa.

Thus, restricting the ontology to 3+1 spacetime creates consistency problems that Cramer has not been successful in resolving. For example, in section 5.7 of his (2016), Cramer addresses the question "Do wave functions exist in real 3D space or only in Hilbert Space?"[8] He considers a photon pair described by a polarization-entangled state $| \Psi >$ which is a superposition of "both photons horizontally polarized" (both H) and "both photon vertically polarized" (both V):

$$| \Psi > = \alpha \, | H_1, H_2 > + \beta \, | V_1, V_2 > \qquad\qquad (2)$$

The entangled photons head towards two separated detectors, each having its own horizontal H and vertical V detectors (indexed by 1 and 2). According to Cramer, each of the individual detectors $H_{1,2}$ and $V_{1,2}$ responds to a 'free, uncorrelated' photon wavefunction that exists in 3-space; thus he appears to deny that the two photons are ever really, physically described by the state given in (2). Instead, he attributes the enforcing of the correlation entirely to the source (along with the conservation laws). Certainly, the conservation laws (along with the transactional process) play a crucial role in enforcing the correlations. But is it not the offer wave itself (2) that specifies what the correlations *are* that need to be enforced?

In any case, it cannot be correct to view the photons as 'uncorrelated' at any stage of the process, given that they were emitted as the entangled two-photon offer wave (2). Subsystems of a composite pure state do not possess individual pure states, correlated or not. In addition, the entire two-photon system is also described by an entangled spatial state; the latter is conventionally not shown when discussing spin-entangled states, but it reflects the fact that the photons have no individuality and cannot really be labeled (1,2). That is, there is not really any fact of the matter about whether 'photon 1' will go to detector 1 or 2, and similarly for 'photon 2'. To say that each detector receives an uncorrelated photon offer wave is to disregard the spatially entangled state that is a crucial component of the overall two-photon total state, and which in this case has 6 spatial degrees of freedom.[9]

---

[8] I would disagree with this formulation of the issue: of course, wave functions (more generally, state vectors) are elements of Hilbert Space. The question is whether *the entities they repres*ent can consistently be considered to be spacetime objects; i.e. whether they can be accurately and consistently represented on a spacetime diagram.

[9] In addition, the probabilities shown by Cramer on p. 72 do not appear to be correct. If confirmations are generated, then the von Neumann 'Process 1' transition is triggered, and the emitted pure state is transformed into a mixed state (sum of weighted projection operators, see for example Kastner 2012, Chapter 3). At this point (assuming an ideal measurement), we will definitely find one of the eigenvalues of the measured observable; there is zero probability of 'no outcome.'

Despite his attempt to preserve a simple spacetime picture, Cramer apparently recognizes that possibilities must be involved. He remarks on p. 72 that "All of the waves involved are physically present, and *in some sense "real," at least at the level of possibility*, in normal three-dimensional space."[10] This ambiguous remark evinces some awareness of the problem of describing offer waves as spacetime objects (given that they represent multiple possibilities, not all of which can be realized in spacetime), but falls far short of addressing the problem. Later on, in Chapter 6, he says that 'reality is not fixed by the initial offer wave' (p. 126) and in Chapter 9, that the future is one of 'multiple possibilities' (p. 165); and that the present emerges from that set of possibilities by way of the transactional process. That ontology is essentially equivalent to the possibilist version of TI (i.e., PTI), discussed in Part I. Thus, there is a pervasive equivocation in Cramer's attempt to restrict his ontology to 3+1 spacetime.

When discussing all transactional opportunities inherent in a single given experimental setup, Cramer comments that only certain transactions can be 'projected out'. But the question arises: projected out of what? One does not project from 3+1 spacetime onto 3+1 spacetime. Projection of many solutions onto few or one is always an operation from a higher dimensional space to a lower one. One cannot have in a single spacetime many different offer wave components leading to many different possible futures. It therefore seems that Kastner (the present author) and Cramer have essentially the same ontology in mind, but that the latter wishes to hold onto a traditional realist description of the processes as if they are occurring in ordinary spacetime, despite the fact that even according to him, there are multiple possibilities in play. This is probably due to the usual tacit assumption that in order to be 'real', something must be a 'spacetime object'. This is a metaphysical view that need not be accepted; it is rejected in PTI.[11]

An expanded, possibilist ontology, despite the challenge it offers to our physical intuitions, should not be taken as problematic or disqualifying for the interpretation. To avoid a double standard, it should be remembered that the DeBroglie-Bohm theory brings with it an expanded ontology whose 'guiding wave' is not a spacetime object, and that is now viewed as a mainstream interpretation. In addition, the currently mainstream 'many worlds' interpretation, in which spacetime splits innumerably and in infinitesimal time intervals, arguably possesses a far more extravagant ontology. In view of the Hilbert Space structure of quantum theory, a thoroughgoing realist interpretation of quantum theory must somehow reflect this higher-dimensional mathematical structure.

In his discussion of situations involving higher-dimensional entangled states, Cramer seems to present TI as a calculational method that is 'applied' to 'vertices' (i.e. at emitters and absorbers). But this is a reversion to instrumentalism about the offers and confirmations that are the primary physical entities in TI. Emitters really emit offer waves, and absorbers really generate confirmations, both of which really are described by their quantum states--however multi-dimensional they might be. An emitted electromagnetic field really exists, whether it's a non-classical entangled state like (2) or a classical field. Cramer is effectively being instrumentalist about the former in his attempt to preserve a 3+1 spacetime ontology.[12]

---

[10] Further ambiguous language about 'quantum possibilities' doing things (but apparently not really existing in the single spacetime manifold required by Cramer's ontology) surfaces in the Question/Answer section (p. 184, questions 1b and 2b).

Thus, Cramer is faced with a dilemma: if all the processes are wholly in spacetime, then he has a block world. In that picture, the dynamical interplay of offers and confirmations is a superfluous story that does not describe anything real, since all the spacetime events are already there. On the other hand, if there is true indeterminism, as Cramer says he intends, then real possibilities are a crucial part of the ontology that cannot be disregarded. The ontology is *de facto* bigger than 3+1 spacetime, and the natural interpretation is that this dynamic manifold of possibility is what is described by the Hilbert Space structure. This is not 'mistaking the map for the territory,' as Cramer asserts (p. 167)[13]; it is acknowledging that, in a realist interpretation, the territory must be bigger than we thought.

Cramer's p. 72 note expresses concern that the possibilist ontology constitutes an 'unnecessary roadblock to visualization'. But it *is* necessary if TI is to be a realist interpretation; and visualization (in the classical sense) is surely not the criterion for correctness of an interpretation. Why should one expect to be able to classically visualize everything that is going on at the quantum level?[14]

Conclusion

This article has reviewed the essential concepts of the Transactional Interpretation of Quantum Mechanics (TI) and its relativistic extension, PTI. According to PTI, collapse is not a process that occurs within spacetime, and that is why it has been so notoriously difficult to give a spacetime account of collapse (cf. Aharonov and Albert (1981) and Kastner (2012, §6.7). Rather, collapse corresponds to the creation of spacetime events from a quantum substratum. That substratum comprises physical possibilities described by quantum states; as well as virtual processes, described by time-symmetric propagators, which are the precursors to those states. Two spacetime events are created via an actualized transaction: (i) the emission and (ii) the absorption of real energy (in the relativistic description, real four-momentum). The connection between these two events is the transfer of real energy/momentum from the emitter to the receiving absorber. The transfer defines a spacetime interval and a temporal direction, the emission defining the past and the absorption defining the present for that absorber. Thus we gain deep physical meaning corresponding to the mathematical facts that energy and momentum are the generators of temporal and spatial translation, respectively.

---

[12] On p. 187 in his (2016), Cramer wonders whether 'instrumentalist' is a 'philosopher's insult'. Of course, 'instrumentalism' about a theory just means that the theory is nothing more than a mathematical formula to calculate results and that no ontology should accompany the theory; it is a form of antirealism. But this appears to be what Cramer is doing with the entangled multiparticle states.

[13] Ironically, here Cramer appropriates a phrase from Kastner (2012, Chapter 2) used therein to argue that quantum theory is a theoretical 'map' which refers to a real ontological territory; this is the opposite of his stated position.

[14] This point was made by McMullin (1984, p. 14).

At the relativistic level, PTI takes into account virtual processes and their couplings with candidate emitters and absorbers. Such virtual processes are necessary but not sufficient conditions for the generation of offers and confirmations. The latter occur on a stochastic basis, where the probabilities for their occurrence correspond to decay rates. Thus, while the transactional picture of measurement involves objective uncertainty, that uncertainty is precisely quantifiable both at the nonrelativistic and relativistic levels.

In addition, it has been argued that a consistent realist ontology for the transactional picture is only achieved by taking the higher-dimensional multi-particle states as referring to real entities that exist in an extra-spatiotemporal manifold describable by Hilbert Space. Assuming an ontology restricted to 3+1 spacetime renders such multiparticle offers and confirmations as instrumentalist constructs, which is contrary to the original realist intent of the interpretation, and which apparently leads to inconsistencies such as describing entangled component systems as uncorrelated.

Acknowledgments.

I would like to thank an anonymous reviewer for valuable suggestions for improvement of the presentation.